\documentclass[revtex4-1,twocolumn,superscriptaddress,apj]{emulateapj}

\newcommand{\erg}{\mathrm{erg}}
\newcommand{\md}{\dot{m}_{\rm Edd}}
\newcommand{\K}{\mathrm{K}}
\newcommand{\gcmthreen}{\mathrm{g}\,\mathrm{cm^{-3}}}
\newcommand{\gcmtwo}{\mathrm{g}\,\mathrm{cm^{-2}}}
\newcommand{\mevu}{\mathrm{MeV}\,\mathrm{u^{-1}}}
\newcommand{\kevu}{\mathrm{keV}\,\mathrm{u^{-1}}}

\usepackage{graphicx}

\begin{document}


\title{Carbon Synthesis in Steady-State Hydrogen and Helium Burning On Accreting Neutron Stars}


\author{Jeremy Stevens\altaffilmark{1,2,3}
Edward F. Brown\altaffilmark{1,2,3},
Andrew Cumming\altaffilmark{4},
Richard Cyburt\altaffilmark{1,2,3},
Hendrik Schatz\altaffilmark{1,2,3}}
\affiliation{1 National Superconducting Cyclotron Laboratory,
Michigan State University, East Lansing, MI 48824, USA }
\affiliation{2 The Joint Institute for Nuclear Astrophysics, 
Michigan State University, East Lansing, MI 48824, USA }
\affiliation{3 Dept.~of Physics and Astronomy, Michigan State
University, East Lansing, MI 48824, USA }
\affiliation{4 Dept.~of Physics, McGill University,Montreal, 
QC H3A 2T8, Canada }


\begin{abstract}
  Superbursts from accreting neutron stars probe nuclear reactions 
at extreme densities ($\rho \approx 10^{9}~\gcmthreen$) and 
temperatures ($T>10^9~\K$). These bursts ($\sim$1000 times more energetic 
than type I X-ray bursts) are most likely triggered by unstable ignition of carbon in a 
sea of heavy nuclei made during the rp-process of regular type I X-ray 
bursts (where the accumulated hydrogen and helium are burned). 
An open question is the origin of sufficient amounts of carbon, which 
is largely destroyed during the rp-process in X-ray bursts. We explore 
carbon production in steady-state burning via the rp-process, which might 
occur together with unstable burning in systems showing superbursts. 
We find that for a wide range of accretion rates and accreted helium 
mass fractions large amounts of 
carbon are produced, even for systems that accrete solar composition. This makes 
stable hydrogen and helium burning a viable source of carbon to trigger superbursts. 
We also 
investigate the sensitivity of the results to nuclear reactions. We find 
that the $^{14}$O($\alpha$,p)$^{17}$F reaction rate introduces by far the 
largest uncertainties in the $^{12}$C yield. 
\end{abstract}



\section{Introduction}
Type I X-ray bursts are thermonuclear flashes of accumulated
hydrogen and helium on the surface of an accreting neutron star
\citep{Woosley1976,Maraschi1977,Joss1977}. Nearly
a hundred Low Mass X-Ray Binaries (LMXBs) in the Galaxy have shown these
events \citep{Galloway2010}, see also the reviews by \citet{Lewin93,Lewin97,
Bildsten1998,Schatz2006a,Parikh2013}.

A rare class of type I bursts, now called superbursts,
were discovered during long term monitoring of LMXBs with the BeppoSAX
Wide Field Cameras \citep{Cornelisse2000,Cornelisse2002,Kuulkers2002}. 
These superbursts occur in sources that otherwise exhibit 
normal type I bursts. Superbursts last about a factor of 1000 longer 
than regular bursts (a day compared to 10--100~s) and have
a factor of 1000 more energy output, typically $10^{42}~\erg$. 
So far about 22 bursts from 13 sources have been observed
(see \citet{Keek2008,Keek2011,Keek2012}).
Recurrence times around a year are estimated but uncertain because of limited observational data
\citep{Keek2006}.  

Superbursts are thought to be driven by thermonuclear ignition of 
a deep carbon layer in the liquid ocean of the neutron star \citep{Cumming2001,Strohmayer2002}. 
The disintegration of heavy rapid proton capture process (rp-process) ashes can provide up to half of the
observed energy \citep{Schatz2003}. 
Because superbursts ignite very close to the outer neutron star crust, they
offer the opportunity to probe crust properties such as thermal structure and conductivity 
\citep{Brown2004,Cumming2006} or physics at the interface between the 
liquid ocean and the solid crust \citep{Horowitz2007,Medin2011}. \citet{Keek2011} demonstrated
the sensitivity of superburst light curves and recurrence times on 
the heat flux emerging from the outer crust, through its effect on the ignition depth. 

The carbon ignition model explains nicely the energetics and the long burst duration that corresponds
to the thermal timescale for cooling such a thick carbon enriched layer. 
By comparing observed light curves to superburst models, \citet{Cumming2004}
and \citet{Cumming2006} obtain ignition column depths of 0.5-3$\times 10^{12}~\gcmtwo$. 
For typical accretion rates, this is consistent with the estimated recurrence times
of the order of a year. Complete superburst light curves from ignition to late time cooling have been modeled
 by \citet{Weinberg2007}. More recently, first multi-zone calculations 
of superbursts have been carried out \citep{Keek2011}. The models
agree reasonably well with observed burst light curves but only when 
using larger than observed accretion rates.

There are also a number of problems with the carbon ignition model. Besides 
insufficient heating leading to too long recurrence times \citep{Keek2008a,Kuulkers2010}, 
the most obvious open question is the production mechanism for the large 
amounts of carbon (mass fraction $X_{\rm C}= $10--20\,\%) needed to ignite the superbursts \citep{Cumming2006}. 
Understanding the origin of the carbon is important to support the carbon ignition model as an 
explanation for superbursts in general, and to constrain the carbon fraction when modeling 
superbursts for comparison with observations. In this work we use 
updated nuclear reaction rates and explore systematically carbon production in stable, steady 
state nuclear burning for a much wider range of parameters that covers not only mixed hydrogen 
and helium accretors, but also helium accreting UCXBs. In addition we investigate the sensitivity 
of carbon production to the underlying nuclear physics for the entire parameter space and 
identify nuclear reaction rates that should be better constrained to reliably predict carbon 
production for superbursts.

\section{Carbon Production in rp-Process Burning}

Most superbursts occur on neutron stars that accrete a mix of hydrogen and helium 
at a typical rate of  0.1--0.3~$\md$ \citep{Falanga2008}, with $\md$
being the Eddington accretion rate for solar composition $\md \approx 10^5~\gcmtwo$\,s$^{-1}$.
In this case carbon has to be produced by thermonuclear burning of hydrogen and helium 
near the neutron star surface. One option are the regular type I 
bursts that these systems exhibit. However, state of the art X-ray burst models  produce 
at most a few percent $^{12}$C \citep{Schatz2003NPA,Woosley2004a,Fisker2008,Jose2010}.
At the end of the burst the ashes contains a mix of  $^{12}$C and unburned $^{4}$He. 
\citet{Woosley2004a} showed that $^{12}$C+$^4$He reactions triggered by heating from subsequent bursts 
on top of the ashes further reduce $X_{\rm C}$ to negligible values.

\citet{Medin2011} recently suggested that in 
steady-state the preferential crystallization of heavy nuclei at the ocean crust interface 
\citep{Horowitz2007} might lead to mixing and an enrichment of light nuclei in the liquid 
phase. The results imply that material with $X_{\rm C}$ of just a few percent could be enriched 
to as much as 40\% at the depth where superbursts ignite. However, the total amount of carbon 
needed to explain the energetics of the observed superbursts remains the same. 
If carbon production is a factor of 10 lower then the time to accumulate the 
required amount of carbon will increase by a factor of 10, making recurrence times
longer than observed. 

Some superburst have been observed in X-ray bursting ultra compact X-ray binaries (UCXB), for 
example in 4U 0614+91 \citep{Kuulkers2010} and 4U 1820-30 \citep{Strohmayer2002}. 
These systems are characterized by very low orbital periods that imply a compact companion star without 
a hydrogen envelope. Most neutron stars in UCXBs are therefore thought to accrete mainly 
$^{4}$He. One might expect that helium driven X-ray bursts are a more favorable environment for 
carbon production due to the low $^{12}$C($\alpha$,$\gamma$) reaction rate creating a bottleneck in the helium 
burning reaction sequence. However, \citet{Weinberg2006} showed that $^{12}$C($\alpha$,p)$^{15}$N(p,$\gamma$)$^{16}$O,
enabled by a very low abundance of hydrogen produced by ($\alpha$,p) reactions, serves as an 
effective bypass of the $^{12}$C($\alpha$,$\gamma$) reaction, resulting in negligible amounts of 
$^{12}$C ($X_{\rm C} < 10^{-5}$) being produced in helium burning X-ray bursts. $^{4}$He driven X-ray burst can also occur 
in mixed hydrogen and helium accretors at low accretion rates and high metallicities, when hydrogen 
burning via the CNO cycle prior to burst ignition consumes  all hydrogen. The environment is 
somewhat different to pure helium accretors as mixing with hydrogen-rich surface layers alters the 
nucleosynthesis. \citet{Woosley2004a} find a carbon production of at most a few precent for such bursts. 

We conclude that neither mixed hydrogen and helium bursts, nor pure helium flashes can produce the 
amounts of carbon required for superbursts. \citet{Kuulkers2010} explored the possibility that 
the explosion of a cold, thick helium layer is responsible for a superburst observed in the UCXB
4U 0614+091. While this model can explain the observed features it would require a nuclear energy 
generation rate of only $0.6\,\mevu$, almost a factor 3 lower than what is expected from helium burning. 
In addition, the model requires atypically low accretion rates around $0.004~\md$ and is 
therefore only applicable for the special case of 4U 0614+091. \citet{Cooper2009} explored
alternative triggers for superbursts, including temperature sensitive electron capture rates and 
$^{12}$C+$^4$He reactions. They find that none of these can explain superburst observables, and conclude that 
deep carbon burning is the only viable model. 

It has been suggested that stable burning of hydrogen or helium is the source of carbon
powering superbursts \citep{Strohmayer2002,Schatz2003NPA}. While
all superbursting systems show regular type I X-ray bursts this does not necessarily mean that 
all the accreted fuel is burned explosively in bursts. \citet{intZand2003} demonstrated that for
all seven superbursting systems they investigated the observed ratio of the bolometric persistent X-ray emission, 
powered by gravitational energy release, and the bolometric summed X-ray emission of all X-ray bursts, 
powered by nuclear burning, is high, of the order of $\alpha \approx 1000$. Such high $\alpha$ values 
are typical for the mixed hydrogen and helium accretors with accretion rates around 0.1--0.3~$\md$ that 
host the majority of superbursts. If all accreted fuel
were burned in X-ray bursts, one would expect $\alpha \approx 40$ --- the ratio of gravitational energy release 
(about $200\,\mevu$) and nuclear energy release (about $5\,\mevu$ for hydrogen burning). Such 
$\alpha$ values are typical for systems that show regular X-ray bursts only. The high $\alpha$ values 
in superbursting systems therefore indicate that a large fraction of the accreted hydrogen and helium is burned stably. 

For the helium accreting UCXB 4U 1820-30 \citet{Cumming2003} find $\alpha=150$, not too far from  what is expected
for helium burning. However, the system is known to spend a significant amount of time in a high accretion 
rate state where bursts disappear. It is therefore possible, that the carbon powering the superburst 
observed during the low accretion rate phase has been formed mainly during stable burning in 
the high accretion rate phase \citep{Strohmayer2002}. 

While observations clearly indicate some stable burning in the accretion rates regime around 0.1--0.3~$\md$
relevant for super bursts, theoretical considerations \citep{Fujimoto81,Bildsten1998} and time dependent models 
of accreting neutron stars \citep{Fisker2007a,Heger2007} indicate a transition to 
stable burning at much higher accretion rates, around $1~\md$. The origin of this discrepancy
is not understood. \citet{Schatz1999} calculated the nucleosynthesis during stable burning of hydrogen and 
helium in solar proportions using a steady-state burning model that assumes the burning to be stable. Calculations 
were performed for a range of accretion rates above $1~\md$ and $X_{\rm C}$ was found to be
at most 6\,\%. Only the $^{15}$O($\alpha$,$\gamma$) reaction rate was varied to explore nuclear 
physics uncertainties. The parameter range was extended to lower accretion rates in 
\citet{Schatz2003NPA}, who found $X_{\rm C}>10\,\%$ for accretion rates below $0.3~\md$. This confirmed 
that steady-state burning is a viable mechanism for producing larger amounts of carbon on 
accreting neutron stars at the typical accretion rates of superburst systems. 


\section{Calculations of Stable rp-Process Burning}
\label{sec:Method}

We use the same model described in \citet{Schatz1999} to 
calculate the nucleosynthesis in steady-state nuclear burning 
on an accreting neutron star. The model determines the 
temperature profile by integrating the 
radiation flux generated by nuclear reactions from the surface down to 
a depth where all fuel is burned. Nuclear reactions are followed with 
a detailed nuclear reaction network including 686 nuclei from hydrogen to 
tellurium using the reactions in JINA reaclib version 1.0 rates \citep{Cyburt2010}. 

The surface gravity is fixed to a typical value of $g=2.42 \times 10^{14}~\gcmtwo$ 
for a canonical neutron star with a mass of 1.4~M$_\odot$ and a 10~km radius, taking 
into account relativistic corrections. This leaves three free parameters, the heat 
flux entering the surface layers from the crust $F_{\rm b}$, the accretion rate $\dot{m}$, 
and the composition of the accreted material, in particular the helium mass fraction 
$X_{\rm He}$. 

$F_{\rm b}$ is the portion of the heat generated by electron captures and pycnonuclear 
fusion reactions deeper in the crust that emerges at the surface of the crust. $F_{\rm b}$ 
depends on the heating and cooling reactions in the crust and the crust's thermal 
conductivity. We explore a typical range of  $F_{\rm b} =$ 1, 50, 100, 500, and 1000 
~$\kevu$ \citep{Brown2000} similar to the range used in studies of superbursts \citep{Keek2011}. 

\begin{figure}[htbp]
\epsscale{1}
\plotone{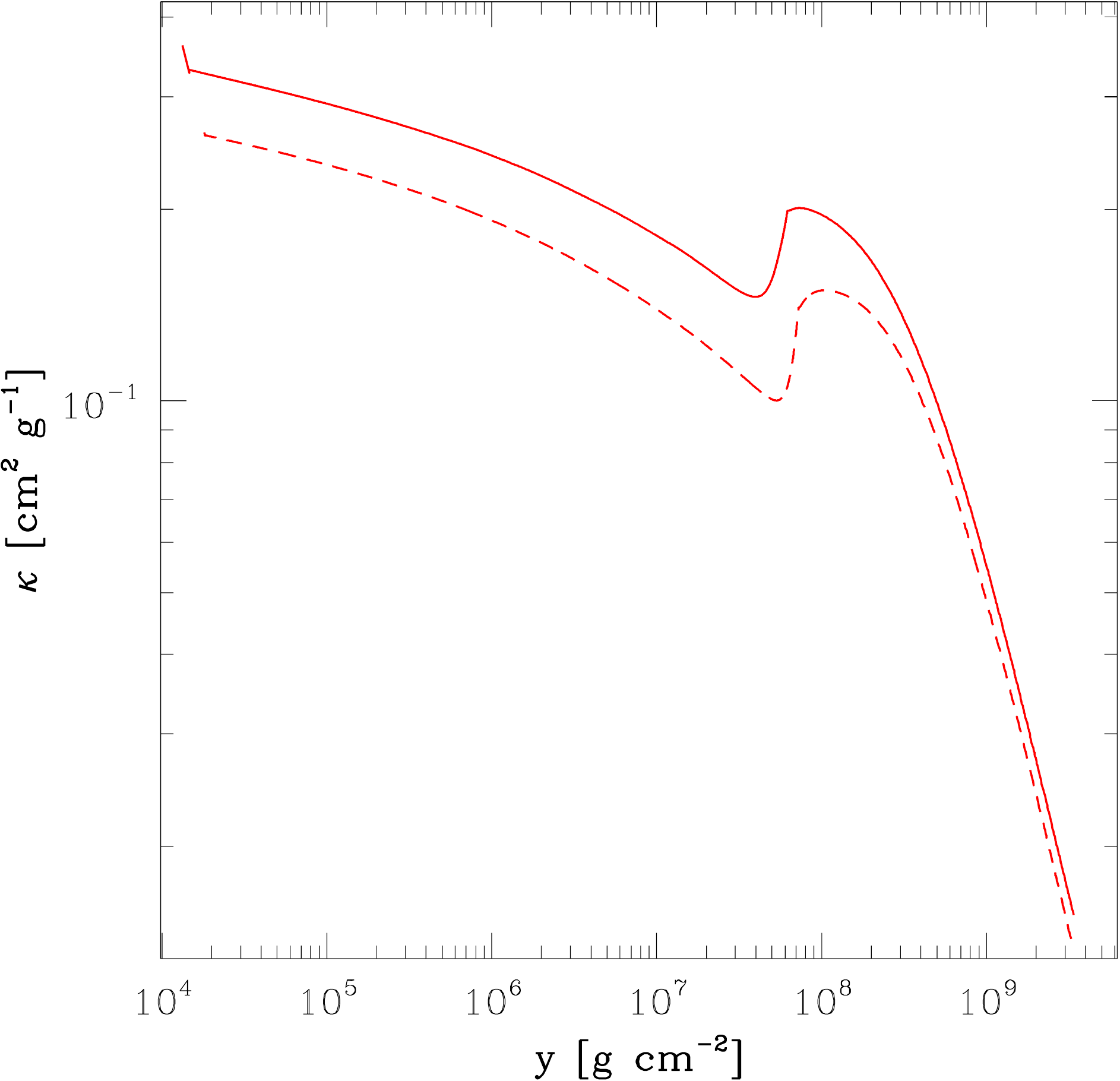}
\caption{The opacity as a function of column depth at $X_{\rm He} = 0.5$ and $\dot{m} = 0.3~\md$ using 
the opacities from \citet{Schatz1999} with (solid) and without (dashed) the non-additivity factor 
from \citet{Potekhin2001}.}
\label{fig:f1}
\end{figure}

The electron scattering, free-free absorption, and conduction opacity calculations used 
in this study are detailed in \citet{Schatz1999}. However, \citet{Schatz1999}
assumed that the total Rosseland mean radiative opacity was given by 
the sum of the electron scattering and free-free Rosseland mean opacities $\kappa=\kappa_{\rm es}+\kappa_{\rm ff}$. 
In fact, as discussed by \citet{Potekhin2001}, the Rosseland means do not add in this 
way, as the Rosseland mean of the opacity sum is not the same as the sum of the Rosseland 
means. A direct sum of the Rosseland means underestimates the opacity by about 30\,\% when 
the electron scattering and free-free contributions are comparable. To correct for this, 
we adopt the ``non-additivity'' factor of \citet{Potekhin2001} (see their eq.~[20]). The new 
opacity is slightly higher (Fig.~\ref{fig:f1}), however, the results are essentially the same
as the ones obtained with the opacities from  \citet{Schatz1999}. 

We express $\dot{m}$ in multiples of $\md=87900$~$\gcmtwo$\,s$^{-1}$, 
which is the local Newtonian Eddington accretion rate for solar composition, and 
provides a convenient scale. 
We vary $\dot m$ from 0.1 to $40~\md$. $\dot m=0.1$--$0.3~\md$ are typical 
values for superbursting systems. However, these accretion rates are derived from 
the inferred global accretion rate assuming the local accretion rate is the same 
across the neutron star surface. This might not always be the case. The fact that 
X-ray burst models only show stable burning at accretion rates significantly above 
$\dot m=0.1$--$0.3~\md$ \citep{Bildsten1998,Fisker2003} might be a hint 
that local accretion rates are higher than expected. To cover this possibility we 
therefore explore also the consequences of higher local accretion rates. 

Accretion from normal main sequence stars likely leads to an accreted composition
close to solar. The composition accreted from a compact helium star in a UCXB is 
expected to be dominated by  $^4$He. Depending on the 
evolutionary path that formed the system, a hydrogen mass fraction between 
close to zero and 40\,\% \citep{Cumming2003}
is possible. To explore the entire parameter space we use solar metallicity throughout (a
test run with reduced metallicity did not show any significant differences)
and vary the hydrogen mass fraction 
from solar to zero. This results in a $X_{\rm He}$ range of 0.28--0.98. 

To explore the sensitivity of 
carbon production to nuclear reaction rates, we vary individual reaction rates
that lie along the reaction path one by one by factors of 10 up and down. The rate 
variations are performed for a range of models on a reduced grid of input parameters. 
We kept $F_{\rm b}=1~\kevu$ as $X_{\rm C}$ was found to not depend significantly on  
$F_{\rm b}$. For $X_{\rm He}$ we chose a low and a high value, 0.3, and 0.9. 
We then used as models for the rate variations $\dot m=0.1, 0.5, 1, 2, 5 
~\md$ and $\dot m=0.1, 0.5, 1, 2, 5, 10, 20, 30~\md$ for low and high $X_{\rm He}$, respectively. 
For low $X_{\rm He}$ a more limited $\dot m$ range is used as carbon production 
is negligible for  $\dot m > 5~\md$.


\section{Carbon Synthesis in Steady-State rp-Process Burning} 
\label{sec:Results}

\begin{figure}[htbp]
\epsscale{1}
\plotone{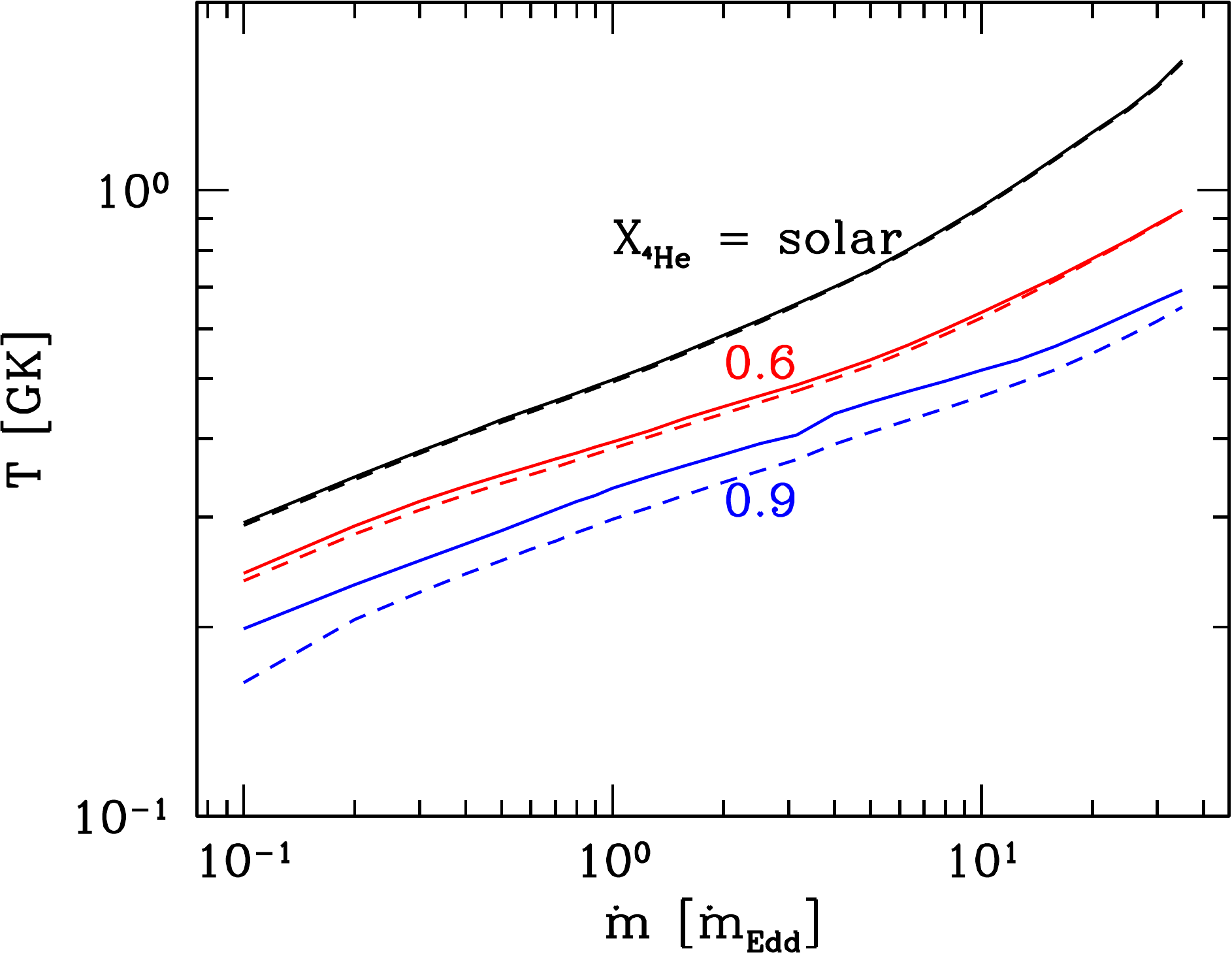}
\caption{Temperature as a function of accretion rate from the time H is reduced to 90\,\% of its 
initial abundance to the time carbon is built up to 90\,\% of its final abundance for 
$X_{\rm He}= $solar(black), 0.6(red), 0.9(blue). Solid is the temperature at $90\,\%$ carbon and dashed is 
the temperature at $90\,\%$ hydrogen.}
\label{fig:f2}
\end{figure}

\citet{Schatz1999} already described the reaction sequences 
during steady-state burning of a solar hydrogen and helium mixture. 
Helium burns mainly by the 3$\alpha$ reaction 
and the $\alpha$p process \citep{Wallace1981}, a sequence of ($\alpha$,p) and (p,$\gamma$) reactions. 
The endpoint of the $\alpha$p process depends very sensitively on peak 
temperature and therefore on accretion rate. Higher accretion rates result in 
a higher nuclear energy production rate generating 
more flux and therefore require a steeper temperature profile resulting 
in a higher burning temperature (Fig.~\ref{fig:f2}). Hydrogen burns at low accretion rates via the CNO cycles 
and at higher accretion rates via the rapid proton capture process  (rp-process) \citep{Wallace1981}, 
a sequence of rapid proton captures and slower $\beta^+$ decays. 

Carbon is produced directly by the 3$\alpha$ reaction but is destroyed rapidly
by the $^{12}$C(p,$\gamma$)$^{13}$N reaction
as long as any hydrogen is present. However, the rate of the 3$\alpha$ 
reaction is proportional to $X_{\rm He}^{3}$ and therefore decreases rapidly 
as $^4$He is consumed. Therefore $^4$He burns at late times much slower than hydrogen.  
The main source of carbon in the ashes is then 3$\alpha$ burning
that occurs after the rp-process has consumed all the hydrogen.

The amount of carbon synthesized is determined by several factors.
One important factor is the amount of helium present at the time of hydrogen
exhaustion, which depends mainly on the speed of the rp-process versus
the 3$\alpha$ process. A slower rp-process leads to more helium
burning before hydrogen is exhausted and therefore less remaining
carbon. The rp-process waiting points therefore play a critical role. 
The second factor is the rate of carbon 
destruction through $^{12}$C($\alpha$,$\gamma$)$^{16}$O once hydrogen 
is consumed,
which depends on the temperature and density after hydrogen exhaustion
\citep{Schatz1999}. In addition, any helium producing hydrogen 
burning reaction sequences
such as the CNO or the SnSbTe cycles \citep{Schatz2001} will favor carbon 
synthesis. 

\subsection{Trends in Carbon Production with $\dot{m}, X_{\rm He}, and F_{\rm b}$}

\begin{figure}[htbp]
\centering{%
\includegraphics[width=0.85\linewidth]{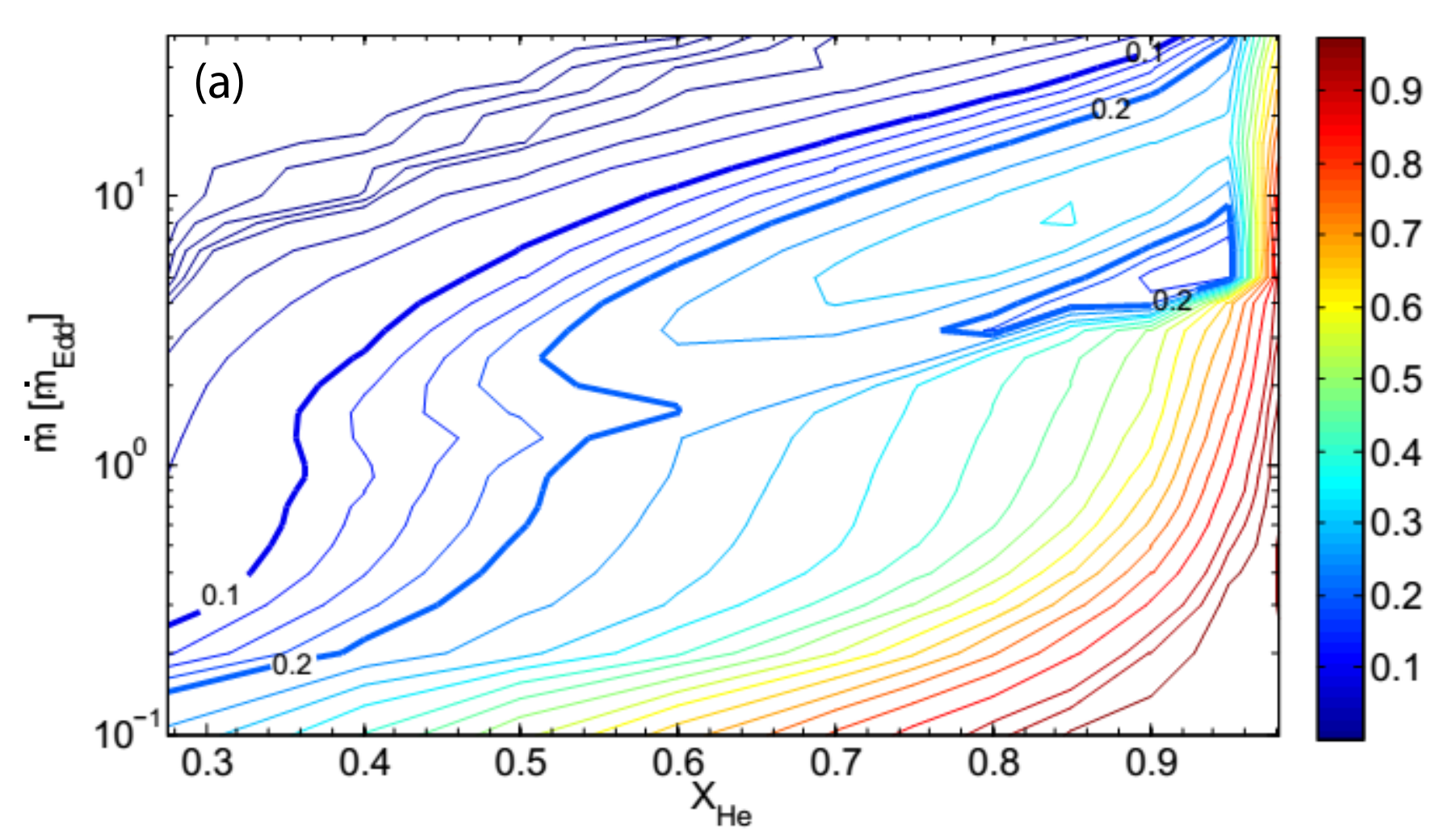}
\includegraphics[width=0.85\linewidth]{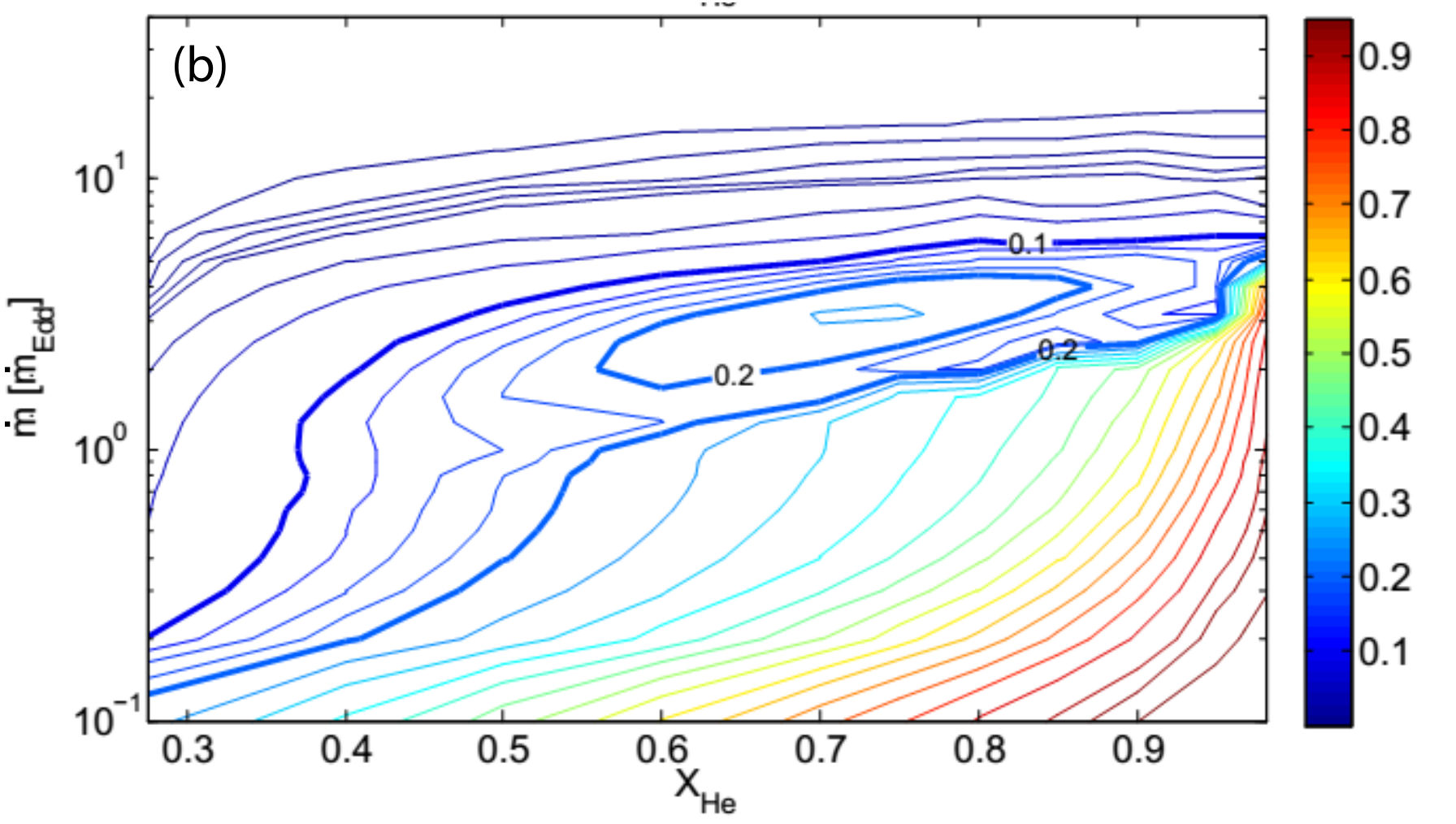}
}
\caption{Final $^{12}$C mass fraction as function of accretion rate and initial 
helium mass fraction at $F_{\rm b} = 1~\kevu$ (a) and $F_{\rm b} = 1~\mevu$ 
(b). Contours are in steps of 0.0025 from 0.001 to 0.01, in steps of 0.025 
from 0.01 to 0.2, in steps of 0.05 from 0.2 to 0.9. Additional contours are drawn 
at 0.925, 0.95, 0.9625, 0.975 and 1. Mass fractions of $10\,\%$ and $20\,\%$ are 
bolded and labeled.}
\label{fig:f3}
\end{figure}

Fig.~\ref{fig:f3} shows the $^{12}$C 
mass fraction produced in steady-state burning at $F_{\rm b}= 1.0~\kevu$. 
There are two overall trends. First,  $X_{\rm C}$ increases with the initial $X_{\rm He}$ 
as higher $X_{\rm He}$ results in more helium being present at the 
point of hydrogen exhaustion. There are a number of reasons for 
this behavior. A higher initial $X_{\rm He}$ simply leads to higher 
$X_{\rm He}$ throughout and therefore to more helium for carbon 
production. Helium burning also generates less energy than hydrogen 
burning. Therefore, a higher initial $X_{\rm He}$ leads to lower 
burning temperatures, and less helium destruction via $^{12}$C($\alpha$,$\gamma$). 
Lower temperatures also lead to a shorter $\alpha$p-process, a lower proton to 
seed ratio, and therefore
a shorter rp-process, which accelerates hydrogen burning
favoring carbon production. 

The second general trend in Fig.~\ref{fig:f3}
is an increasing $X_{\rm C}$  with decreasing $\dot m$. One reason
is that carbon production is favored at lower temperatures (see arguments 
above). In addition, at low $\dot m$ hydrogen burning via the rp-process
sets in at much later times, resulting in more hydrogen burned via the CNO 
cycle prior to ignition of the rp-process, further increasing $X_{\rm He}$.

Carbon production depends only modestly on $F_{\rm b}$ (Fig.~\ref{fig:f3}). 
The biggest difference occurs for $\dot{m} > 10~\md$ where for 
$F_{\rm b}=1~\mevu$ carbon production rapidly goes to zero regardless of composition, while for 
$F_{\rm b}=1~\kevu$ for $X_{\rm He} > 0.6$ still more than 10\% carbon can be made. The reason 
is that for a significant $F_{\rm b}$ the temperature will continue to raise beyond the 
depth of the main nuclear energy generation, leading to more efficient 
burning of the remaining carbon and helium.

\subsection{An Island of Lower Carbon Production}

There is a narrow range of accretion rates --- around $1~\md$
for near solar $X_{\rm He}$ increasing to around 8--10~$\md$ for 
helium-rich environments --- where less carbon is produced than is 
expected from the general trends. For low $X_{\rm He}$ this is just a 
small dip, but for larger $X_{\rm He}$ the reduction in carbon 
production is substantial. This gives rise to an island of low carbon 
mass fraction in the contours of Fig.~\ref{fig:f3}.

\begin{figure}[htbp]
\epsscale{1}
\plotone{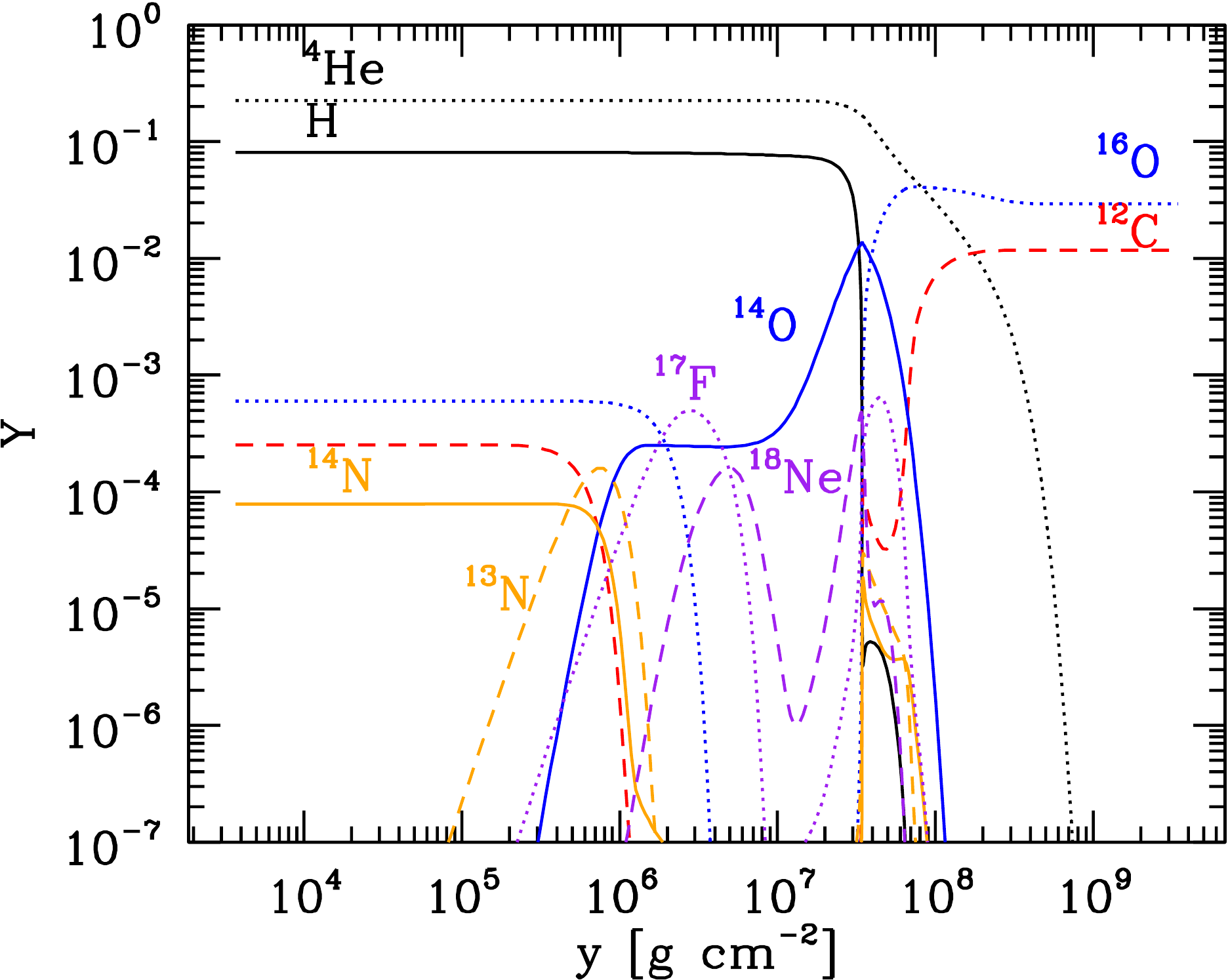}
\caption{Abundances of H, He, $^{12}$C, $^{13}$N, $^{14}$N, $^{14}$O, 
$^{16}$O, $^{17}$F, and $^{18}$Ne as functions of column depth, represented by 
black(s), black(d), red(s), orange(s), orange(d), blue(s), blue
(do), indigo(do), and indigo(d) lines, respectively, with ``s'', ``d'', and ``do''
indicating solid, dashed, and dotted respectively. 
Conditions correspond to the island of low carbon production, 
 $F_{\rm b} = 1~\kevu$ and $\dot{m} = 5~\md$.}
\label{fig:f4}
\end{figure}

The reason for this behavior is the interplay between the
$^{14}$O($\alpha$,p)$^{17}$F reaction and the $^{12}$C(p,$\gamma$)$^{13}$N($\alpha$,p)$^{16}$O
carbon destruction sequence once most of the hydrogen is consumed. The destruction of carbon via 
$^{12}$C(p,$\gamma$)$^{13}$N($\alpha$,p)$^{16}$O requires only a
very small abundance of hydrogen \citep{Weinberg2006}, since hydrogen merely is needed
as a catalyst. For some accretion rates the $^{14}$O($\alpha$,p)$^{17}$F reaction can 
maintain such a small proton abundance during the helium burning phase, 
reducing the amount of carbon that can be produced. 
This can be seen in Fig.~\ref{fig:f4}. After hydrogen burning, as $^{14}$O is depleted, the hydrogen abundance 
starts increasing again reaching abundances around 10$^{-6}$. During this time, the carbon abundance
is significantly reduced. The hydrogen abundance level that can be maintained by the $^{14}$O($\alpha$,p)$^{17}$F
reaction also depends on the rate of the  $^{17}$F(p,$\gamma$) reaction. Once $^{14}$O is depleted, 
$^{17}$F(p,$\gamma$) destroys the remaining hydrogen and carbon production increases again, however, the 
final carbon abundance is significantly reduced. 

The effect can only occur for a narrow range of accretion rates where at the time of hydrogen 
exhaustion a substantial amount of $^{14}$O has build up and starts burning via $^{14}$O($\alpha$,p). 
For lower accretion rates and lower temperatures, $^{14}$O($\alpha$,p) will not occur during 
helium burning, or, $^{14}$O is not produced at all. For higher accretion rates and higher temperatures 
$^{14}$O($\alpha$,p) breakout occurs prior to hydrogen exhaustion destroying all $^{14}$O. In principle 
($\alpha$,p) reactions on heavier nuclei serve then as sources of protons. However, the proton abundance 
level that is reached never exceeds 10$^{-8}$, which is not sufficient to trigger the 
$^{12}$C(p,$\gamma$)$^{13}$N($\alpha$,p)$^{16}$O carbon destruction sequence.

\section{Reaction Rate Sensitivity}

To better understand the dependence of the synthesis of carbon on nuclear reactions, and 
to identify bottleneck reactions  
we varied all reaction rates with significant reaction flow by a factor of 10 up and down for a set of 
accretion rates and initial compositions of $X_{\rm He}=0.9$ (helium-rich environment) and $X_{\rm He}=0.275$ 
(hydrogen-rich environment). All important reaction rates that change the final $^{12}$C mass fraction by more 
than 5$\,\%$ when varied by a factor of 10 are summarized in Table~\ref{tab:t1}, 
along with the maximum change in carbon mass fraction.

\begin{deluxetable}{lcccc}[h]
\tablecaption{Reactions impacting carbon production by more than 5\,\% when varied 
by a factor of 10 for conditions where more than $X_{\rm C}=0.01$ is produced.
\label{tab:t1}}
\tablehead{
\colhead{Reaction} & \colhead{x10\tablenotemark{a}} & \colhead{/10\tablenotemark{b}} & \colhead{Experimentally uncertain rate}
}
\startdata

  $^{12}$C(p,$\gamma$)$^{13}$N & 0.04 & 2.58 &\\
  $^{13}$N(p,$\gamma$)$^{14}$O & 0.77 & 1.12 &\\
  $^{13}$N($\alpha$,p)$^{16}$O & 1.12 & 0.75 & Yes\\
  $^{14}$O($\beta^+)^{14}$N & 3.83 & 0.19 &\\
  $^{14}$O($\alpha$,p)$^{17}$F & 2.70 & 2.79 & Yes\\
  $^{15}$O($\beta^+)^{15}$N & 2.00 & 0.30 &\\
  $^{18}$F($\beta^+)^{18}$O & 0.94 & 1.12 &\\
  $^{18}$Ne($\beta^+)^{18}$F & 1.40 & 0.17 &\\
  $^{19}$F(p,$\alpha$)$^{16}$O & $\approx 1$ & 0.88 &\\
  $^{19}$Ne($\beta^+)^{19}$F & 0.64 & 0.79 &\\
  $^{22}$Na(p,$\gamma$)$^{23}$Mg & 1.54 & 0.77 &\\
  $^{22}$Na($\alpha$,p)$^{25}$Mg & 1.09 & 1.23 & Yes\\
  $^{43}$Ti(p,$\gamma$)$^{44}$V & $\approx 1$ & 1.14 & Yes\\
  $^{52}$Fe(p,$\gamma$)$^{53}$Co & $\approx 1$ & 0.94 & Yes\\
  $^{57}$Ni(p,$\gamma$)$^{58}$Cu & $\approx 1$ & 0.92 & Yes\\
  $^{61}$Ga($\beta^+$)$^{61}$Zn & 1.06 & $\approx 1$ &\\ 
  $^{64}$Ge($\beta^+$)$^{64}$Ga & 1.09 & 0.84 &\\
  $^{66}$Ge(p,$\gamma$)$^{67}$As & $\approx 1$ & 0.95 & Yes\\
  $^{68}$Se($\beta^+$)$^{68}$As & $\approx 1$ & 0.94 &\\
\enddata
\tablenotetext{a}{Factor of change in $X_{\rm C}$ when increasing rate by a factor of 10}
\tablenotetext{b}{Factor of change in $X_{\rm C}$ when decreasing rate by a factor of 10}
\end{deluxetable}

\begin{figure*}[tp]
\includegraphics[width=\linewidth]{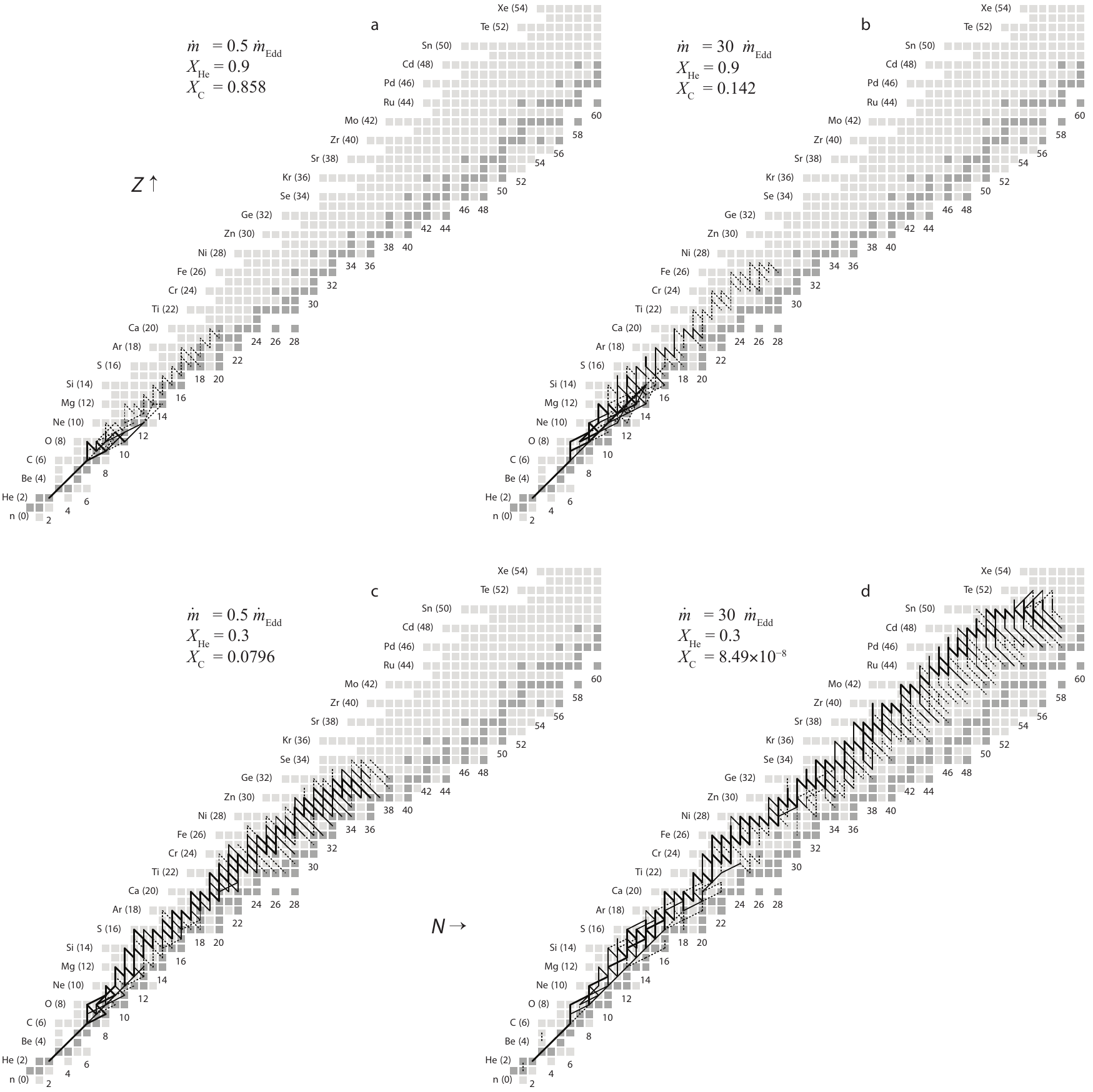}
\caption{Time-integrated reaction flows for (a) $X_{\rm He} = 0.9$, $\dot{m} = 0.5~\md$; (b) $X_{\rm He} = 0.9$, $\dot{m} = 30~\md$; (c) $X_{\rm He} = 0.3$, $\dot{m} = 0.5~\md$; and (d)
$X_{\rm He} = 0.3$, $\dot{m} = 30~\md$.  The final $^{12}\mathrm{C}$ mass fractions for these runs are $X_{\mathrm{C}} = 0.858$, $0.142$, $0.0796$, and $8.49\times 10^{-8}$, respectively.}
\label{fig:f5}
\end{figure*}

At low accretion rates nuclear processing is largely dominated by the hot CNO cycle with the $\beta$-decays of 
$^{14}$O and $^{15}$O being the only reactions controlling carbon production (Fig.~\ref{fig:f5}). At 
somewhat higher accretion rates (starting at 2 and $0.5~\md$ in helium and hydrogen-rich environments, 
respectively) $^{12}$C(p,$\gamma$)  becomes one of the most important reactions as the primary destruction 
mechanism for carbon. 

In helium-rich material the $^{14}$O($\alpha$,p) triggered carbon depletion effect discussed above occurs 
at accretion rates in the $5\textrm{--}20\,\md$ range. Under those conditions the proton production 
reaction rate $^{14}$O($\alpha$,p)$^{17}$F and the rates of the carbon destruction sequence 
$^{12}$C(p,$\gamma$) and $^{13}$N($\alpha$,p) become important. In addition the 
$^{19}$Ne $\beta$-decay and the $^{19}$F(p,$\alpha$) reaction rates are important, as these reactions 
form an additional pathway to produce helium thereby increasing final carbon production. However, the 
impact of the reaction rates in such cycles is not obvious as the impact from the increased helium 
production can be offset by the slower hydrogen burning in the cycle compared to the more rapid 
hydrogen burning reactions sequences of the rp-process. For example, a decreased $^{19}$F(p,$\alpha$) 
rate decreases carbon production, while both, and increased and a decreased $^{19}$Ne $\beta$-decay 
rate decrease carbon production. 

In helium-rich environments at very high accretion rates of $30~\md$
the NeNa cycle \citep{Marion1957} becomes important (Fig.~\ref{fig:f5}), resulting in 
sensitivity of carbon production to the rates of the $^{22}$Na(p,$\gamma$) and $^{22}$Na($\alpha$,p) 
breakout reaction rates.

In hydrogen-rich environments more complex reaction sequences develop (Fig.~\ref{fig:f5}). Already between 
0.1--0.5~$\md$ breakout from the CNO cycles occurs. At 
$0.5~\md$ $^{14}$O($\alpha$,p) becomes important causing the additional carbon depletion effect 
described above. The $^{43}$Ti(p,$\gamma$) reaction rate becomes 
also important determining breakout from the CaSc \citep{VanWormer1994} cycle. 
This cycle slows down hydrogen burning and increases helium production. In addition 
rp-process bottleneck reactions such as $^{57}$Ni(p,$\gamma$) start controlling carbon production. 
For such rp-process bottle-necks, an increased rate leads to more rapid hydrogen burning, earlier 
hydrogen consumption, higher helium abundance at the time of hydrogen depletion, and therefore 
increased carbon production. 

At higher accretion rates of $2~\md$ the important rp-process bottlenecks are the $\beta^+$ decays 
of $^{61}$Ga, $^{64}$Ge, and $^{68}$Se. The sensitivity to the $^{18}$Ne $\beta^+$ decay rate indicates the 
importance of the extended CNO cycle. At still higher accretion rates carbon production is of the order of a 
percent or less and therefore negligible. 

\begin{figure}[htbp]
\centering{
\includegraphics[width=0.85\linewidth]{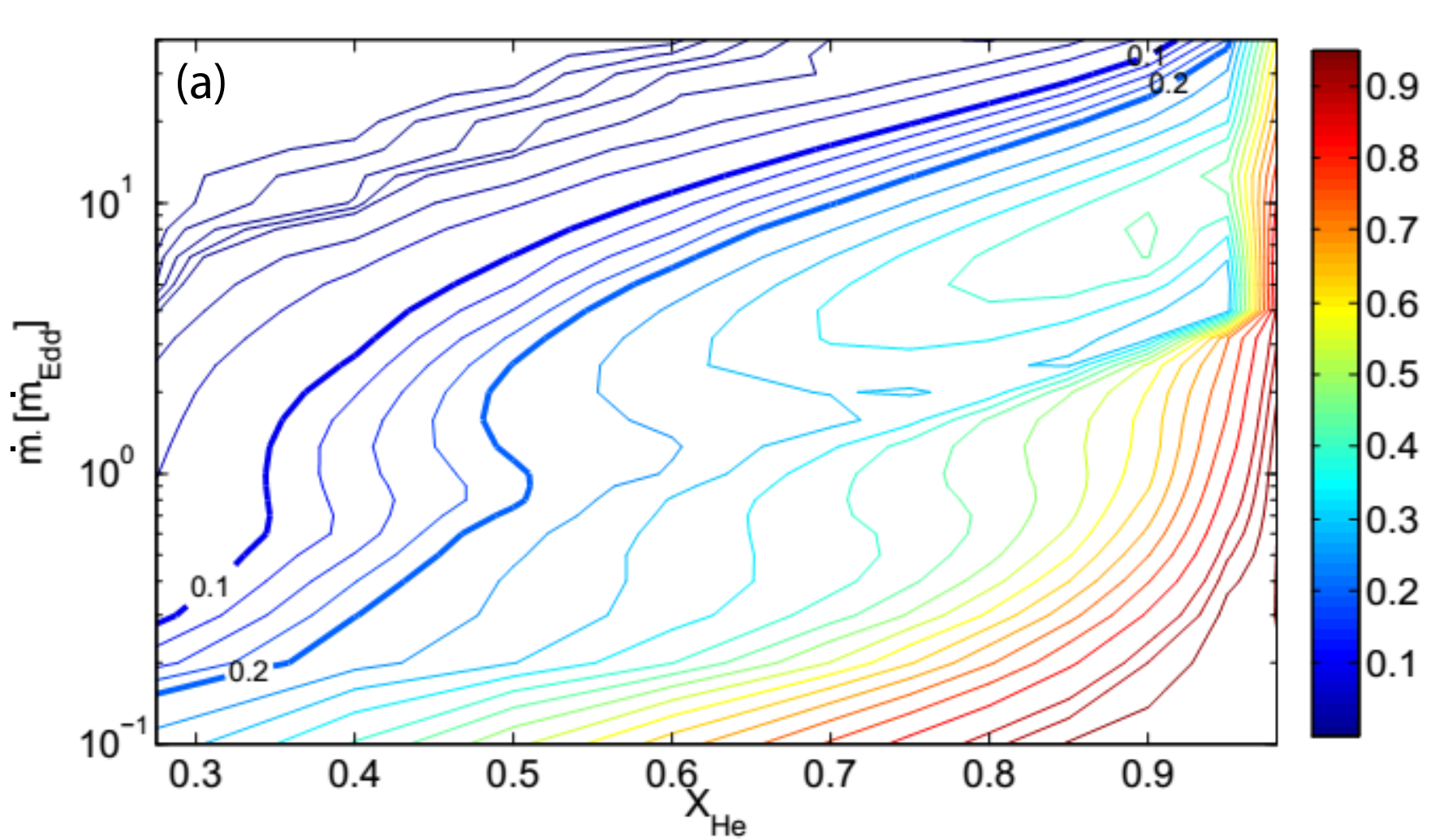}
\includegraphics[width=0.85\linewidth]{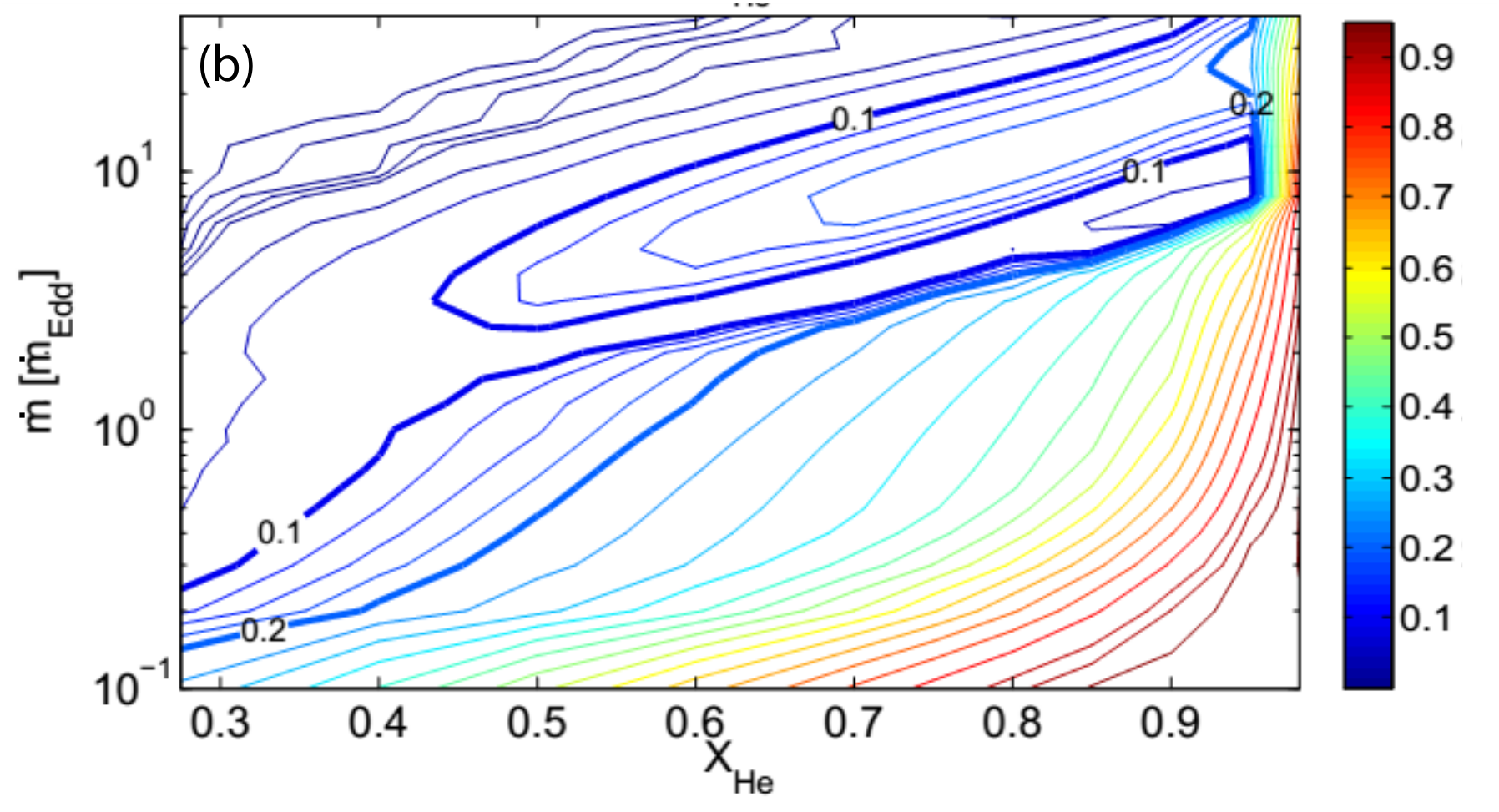}
}
\caption{Final $^{12}$C mass fraction as function of accretion rate and 
initial helium mass fraction at $F_{\rm b} = 1~\kevu$ with the 
$^{14}$O($\alpha$,p)$^{17}$F reaction rate  increased (a) and decreased 
(b) by a factor of 10. See Fig. \ref{fig:f3} for contour details.}
\label{fig:f6}
\end{figure}

\begin{figure}[htbp]
\epsscale{1}
\plotone{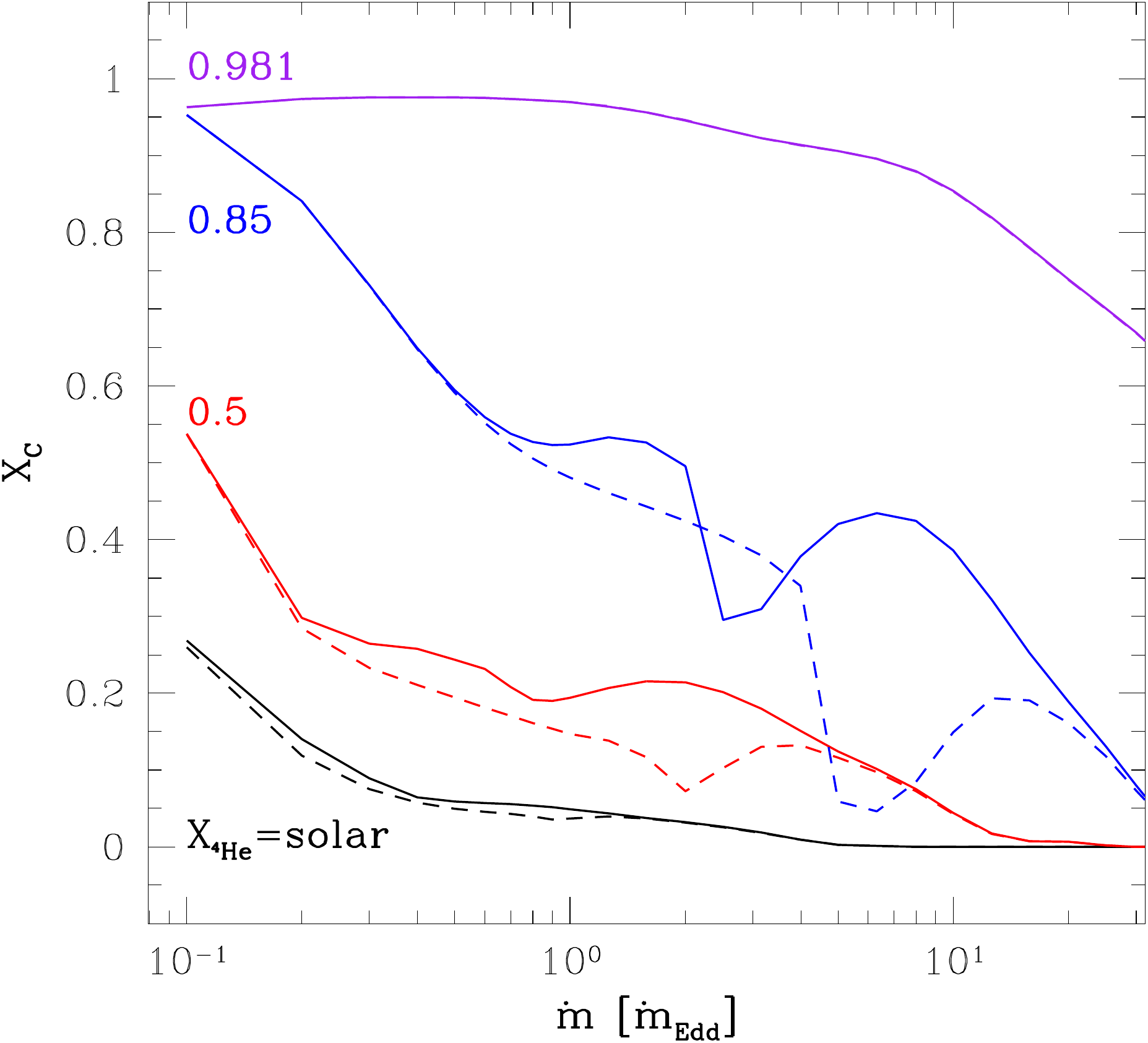}
\caption{Carbon mass fraction $X_{\rm C}$ as a function of accretion rate $\dot{m}$ 
for several initial $X_{\rm He}$  and for the $^{14}$O($\alpha$,p)$^{17}$F reaction rate 
increased by a factor of 10 (solid lines) and decreased by a factor of 10 (dashed lines).
Black, red, blue, and purple lines represent, in order, $X_{\rm He}$ = solar, 0.5, 0.85, and 0.98. 
}
\label{fig:f7}
\end{figure}

The ($\alpha$,p) and ($\beta$+) reactions on $^{14}$O at high He 
mass fractions and around $\dot m$= 5 and $10~\md$ have the 
largest impact on carbon production. The $^{14}$O($\alpha$,p)$^{17}$F reaction has the largest 
impact on carbon production because changes in its rate shift the conditions where the released 
protons can lead to additional carbon destruction. As Fig.~\ref{fig:f6} and \ref{fig:f7} 
show, a higher rate shifts the narrow parameter space for additional carbon destruction to 
lower $\dot m$ as the rate becomes already effective at lower temperatures. At the same
time the strength of the effect is reduced as the higher rate results in a reduced buildup 
of $^{14}$O. For helium-rich environments ($X_{\rm He} > 0.5$) with some small admixture of 
hydrogen a high $^{14}$O($\alpha$,p)$^{17}$F therefore allows for the production
of large amounts of carbon $X_{\rm C} > 20\%$ even at very high accretion rates 
(3--20~$\md$ depending on $X_{\rm He}$) beyond the narrow $^{14}$O($\alpha$,p)$^{17}$F carbon 
destruction region. For a low $^{14}$O($\alpha$,p)$^{17}$F reaction rate carbon production
is much reduced for high accretion rates. 

Except for the narrow region of additional carbon destruction, a large
$^{14}$O($\alpha$,p)$^{17}$F reaction rate therefore increases carbon 
production in most areas of the parameter space. This is also true for 
lower accretion rates below the carbon destruction region, even for 
hydrogen-rich environments. In this regime a higher 
$^{14}$O($\alpha$,p)$^{17}$F rate will reduce the buildup of 
$^{14}$O, and therefore the amount of protons released, once 
hydrogen is consumed. While at the lower temperatures in this 
accretion rate regime the $^{12}$C(p,$\gamma$)$^{13}$N($\alpha$,p)$^{16}$O chain 
cannot be triggered, proton production will still lead to some carbon destruction. 


\section{Conclusions}

We have explored carbon production in steady-state hydrogen and 
helium burning. We find 
that large amounts of carbon $X_{\rm C}  > 10$--$20\,\%$ sufficient for igniting superbursts are produced 
for a wide range of parameters. While carbon production 
increases with $X_{\rm He}$ in the accreted material, we find that 
even for solar composition  $X_{\rm C} > 10\,\%$ for 
$\dot{m} < 0.28~\md$. Therefore, steady-state 
burning is expected to produce enough carbon to power 
superbursts for typical observationally inferred accretion rates, assuming the accretion 
rate at the depth of nuclear burning is the same across the neutron star surface. 

The problem remains that current models do not predict steady-state burning 
at such low accretion rates even though it is observed. One possible solution 
is rotationally induced mixing that transports fuel to greater depths more efficiently
\citep{Keek2009}. Another possibility is that 
 anisotropies in the accretion rate persist to the depth of nuclear burning 
leading to higher local accretion rates in certain regions \citep{Bildsten1998}. 
However, our results 
indicate that in such cases, the accreted material must be depleted in 
hydrogen to produce large amounts of carbon. For example, at a local 
accretion rate of $1~\md$, roughly where models predict the 
transition to stable burning \citep{Bildsten1998},
depending on the value of the $^{14}$O($\alpha$,p) reaction rate, 
$X_{\rm He} > $0.36--0.4 is required to produce $X_{\rm C} > 0.1$. Higher 
local accretion rates will require even higher $X_{\rm He}$. 

We also find that carbon production is reduced for certain accretion rates
due to proton production via the $^{14}$O($\alpha$,p) reaction at late times, once 
hydrogen has been consumed. The effect on carbon production is particularly 
pronounced if the $^{14}$O($\alpha$,p) reaction 
rate is a factor of 10 lower than currently estimated. In this case, 
the only way to produce $X_{\rm C} \approx 20\,\%$ at accretion rates
beyond $10~\md$ is for a pure helium composition. 

We identified the critical nuclear physics that controls carbon production 
in steady-state burning. A quantitative determination of the nuclear 
physics uncertainty in $X_{\rm C}$ would require a Monte Carlo study where
all reaction rates are varied randomly for all conditions. This is beyond
the scope of this study. Nevertheless we can use the results of our sensitivity 
study to identify sources of uncertainty that need to be addressed for a
reliable prediction of carbon production in this scenario. 

Among the reactions listed in Table~\ref{tab:t1} the $\beta^+$ decay rates
are not expected to be major sources of uncertainty. 
All the rates listed are well known experimentally. Terrestrial electron capture contributions 
are expected to be small, and the nuclei do not have low lying excited
states that could be thermally populated in a significant way. Therefore 
modifications to these decay rates due to the high densities and temperatures
in the astrophysical environment are expected to be small. 

The reaction rates for proton capture on 
$^{12}$C, $^{13}$N, and $^{22}$Na as well as $^{19}$F(p,$\alpha$) 
are relatively well experimentally studied with uncertainties of the order of 25\,\% or 
less \citep{NACRE1,Iliadis2010,NACRE2,LaCognata2011}. Considering the impact that a factor of 
10 variation of these rates have on carbon production we conclude that their uncertainties
affect $X_{\rm C}$ by less than 5\,\% and are therefore negligible. 

For the remaining reaction rates experimental information is sparse or non existent 
and assuming an uncertainty of the order of a factor of 10 is not unreasonable.
These reaction rates introduce uncertainties of the order of 5--10\,\% or more in 
$X_{\rm C}$ and are marked in Table~\ref{tab:t1}. Only the uncertainty of the 
$^{14}$O($\alpha$,p) reaction rate affects $X_{\rm C}$ by more than 50\,\%.
It should be noted that for most of these reactions estimates of the uncertainty are 
highly uncertain themselves, and that larger deviations from the true rate are not unlikely. 
A better experimental determination of these reaction rates would significantly 
improve the reliability of predictions of carbon production in steady-state burning of 
hydrogen and helium.


\acknowledgements

We thank L. Bildsten, L. Keek, and Z. Meisel for useful discussions, and F.-K. Thielemann for providing the reaction network solver. 
A. Cumming is supported by an NSERC Discovery Grant and is an Associate Member of 
the CIFAR Cosmology and Gravity program. E. F. Brown is supported NSF AST grant 11-09176. E.~F. Brown and A. Cumming are grateful to the International Space Science 
Institute (ISSI) in Bern for the support of an International Team on Type I X-ray Bursts. This work was supported 
by NSF grants PHY 08-22648 (Joint Institute for Nuclear Astrophysics) and NSF PHY 11-0251.


\bibliography{sspaper_final.bbl}

\end{document}